\documentclass[superscriptaddress,floatfix,prl,twocolumn,amsmath,amssymb]{revtex4-2}

\usepackage{graphicx}
\usepackage{dcolumn}
\usepackage{bm}
\usepackage{amssymb}
\usepackage{amsfonts}
\usepackage{natbib}
\usepackage{amsmath}
\usepackage{mathtools}
\usepackage{xcolor}
\usepackage{color}
\usepackage{datetime}
\usepackage{footnote}
\usepackage{bbold}
\usepackage[normalem]{ulem}
 \usepackage{ragged2e}
 \usepackage{xfrac}
 \usepackage{sidecap}
 \usepackage{setspace}
 \usepackage{multirow}
 \usepackage{xcolor}
 \usepackage{flushend}
\usepackage{hyperref}
\usepackage{cleveref}
\usepackage{microtype}
\usepackage{siunitx}
\usepackage{braket}
\usepackage{float}

\newcommand{\gray}[1] {\textcolor{lightgray}{#1}}
\newcommand{\highlightblue}[1]{%
  \colorbox{blue!20}{$\displaystyle#1$}}
\newcommand{\highlightgreen}[1]{%
  \colorbox{green!20}{$\displaystyle#1$}}
\newcommand{\highlightyellow}[1]{%
  \colorbox{yellow!20}{$\displaystyle#1$}}

\begin{document}

\author{Tim F. Weiss } % e-mail: timweiss001@gmail.com
\address{Quantum Photonics Laboratory and Centre for Quantum Computation and Communication Technology, RMIT University, Melbourne, VIC 3000, Australia}
\email{timweiss001@gmail.com}

\author{Alberto Peruzzo}
\address{Quantum Photonics Laboratory and Centre for Quantum Computation and Communication Technology, RMIT University, Melbourne, VIC 3000, Australia}
\address{Qubit Pharmaceuticals, Advanced Research Department, Paris, France}
\email{alberto.peruzzo@rmit.edu.au}

\title{Nonlinear Domain Engineering for Quantum Technologies}

\begin{abstract}

\bigskip

%Abstract
The continuously growing effort towards developing real-world quantum technological applications has come to demand an increasing amount of flexibility from its respective platforms. This review presents a highly adaptable engineering technique for photonic quantum technologies based on the artificial structuring of the material nonlinearity. This technique, while, in a simple form, already featured across the full breadth of photonic quantum technologies, has undergone significant development over the last decade, now featuring advanced, aperiodic designs. This review gives an introduction to the three-wave-mixing processes lying at the core of this approach, and illustrates, on basis of the underlying quantum-mechanical description, how they can artificially be manipulated to engineer the corresponding photon characteristics. It then describes how this technique can be employed to realize a number of very different objectives which are expected to find application across the full range of photonic quantum technologies, and presents a summary of the research done towards these ends to date.
\end{abstract}

\maketitle

\noindent
\textbf{Introduction}\\
The far-reaching field of quantum technologies has long developed to encompass a large number of vastly different applications, including complex computing architectures \cite{LaddObrian:10,OBrian:07}, long-range distributed networks for the transmission of information \cite{Kimble:08,GisinZBinden:02,DuanZoller:01}, and compact devices leveraging the superior measurement precision available to quantum systems \cite{SlussarenkoPryde:17,DegenCappellaro:17}. Many of these applications have been demonstrated in a range of different platforms.

Among these, photonics-based quantum technologies hold particular promise, featuring a number of unique advantages, both due to their intrinsic suitability for quantum-mechanics-based applications and their technological potential: 

(\textit{1}) Photons can be generated and manipulated in complex, high-dimensional quantum states in several different degrees-of-freedom, and are essentially decoupled from their environment, enabling long-range transmission and room-temperature operation \cite{OBrianVuckovic:09}. 

(\textit{2}) Photons readily interact with other quantum systems, like, for instance, atomic ensembles \cite{ParigiLaurat:15,HammererPolzik:10} or single atoms \cite{ReisererRempe:15}, trapped ions \cite{BruzewiczSage:19}, or solid-state defects \cite{AwschalomZhou:18}, and can be used to connect isolated elements, or form an interface between different platforms. 

(\textit{3}) Generation, control and detection can be realized in an integrated manner, promising exceptional device stability and up-scaling potential \cite{WangThompson:20}.

Due to the vast number of possible, different technologies, real-world implementation and efficient up-scaling efforts require increasing amounts of system flexibility and engineerability, to ensure compatibility between the platform and the different applications. To this end, a wide range of components, based on photonic circuits, mechanical- and electro-optical interaction, or quantum interference effects have been developed to directly manipulate photonic quantum states \cite{WangThompson:20,ElshaariZwiller:20,SlussarenkoPryde:19,LaukSimon:20}. 

Here, we study a complementary engineering approach by addressing the processes governing photon generation, originating from the photons' nonlinear interaction with the surrounding crystal. This allows the creation of photons in specific optical modes and precisely designed quantum states. In particular, this review will focus on approaches relying on the artificial structuring of the material nonlinearity, which, naturally present in a uniform fashion, represents an essentially unused degree-of-freedom in the photon generation process \cite{HuZhu:13,ArieVoloch:10}.

In the following, we will focus on the strongest, second-order nonlinear interaction, governing the processes known as three-wave-mixing, in the single-photon regime, applicable to quantum technologies. These interactions include spontaneous parametric down-conversion (SPDC), which can be thought of as 'spontaneous' down-conversion of a high-energy pump photons into photon pairs of lower energy, generally referred to as signal and idler \cite{Couteau:18}; and single-photon frequency conversion processes, mediated by sum- and difference-frequency-generation (SFG/DFG), in which a signal photon is up-/down-converted to a frequency given by the sum/difference of the signal and the pump frequency \cite{RaymerSrinivasan:12}.

SPDC is commonly used to generate photons of particular quantum properties at specific frequencies, whereas sum- and difference-frequency generation is employed to conserve entanglement and quantum statistics of an input photon while converting it to another wavelength. Both processes require a strong, classical pump field to occur efficiently.

Highly engineerable control of SPDC and sum- or difference-frequency generation can be realized by artificially structuring the material to manipulate its nonlinearity tensor \cite{HumFejer:07}. This is achieved by inversion of locally constrained nonlinear domains, resulting in a nonlinearity either orientated 'up' or 'down', with positive or negative signs respectively, corresponding to the orientation of the optical axis \cite{ShurBaturin:15}. Designed arrangement of these domains - the approach we will refer to as domain engineering - allows to adapt these processes to a number of vastly different applications. Specifically, as we will show later, domain engineering provides the means to directly control two-photon entanglement, polarization and spectral characteristics.

Targeted inversion of nonlinear domains, commonly referred to as 'poling', is achieved, in certain crystals, by locally applying a strong electric field. This can be accomplished via a number of different techniques \cite{YamadaWatanabe:93,LiKitamura:05,MutterCanalias:22,XuZhang:22}. Lithium Niobate (LN) \cite{SaraviSetzpfandt:21} and Potassium Titanyl Phosphate (KTP) \cite{BierleinVanherzeele:89} have been the most prominent host material for such techniques, with inverted domains reported as low as $<1\mathrm{\mu m}$. Both materials can be fashioned into waveguides, enhancing the nonlinear interaction by several orders of magnitude \cite{FiorentinoMunro:07,HeltSipe:12}. LN furthermore features strong electro-optic interaction and is now readily available as thin-film Lithium Niobate on-insulator, constituting a platform fully compatible with integrated photonic components and techniques.

Artificial structuring of the nonlinearity can alternatively be achieved during material growth using so-called orientation pattering, which has been used to demonstrate simple domain engineering in the semiconductors Gallium Arsenide and Gallium Phosphide \cite{SchunemannBudni:16}. The associated fabrication techniques are significantly less matured than those available for the manipulation of LN and KTP, but both materials offer exceptionally strong nonlinear interaction and transparency deep into the mid-infrared.

The analysis presented below will be limited to techniques relying on one-dimensional structuring of the material nonlinearity, as designs relying on higher-dimensional manipulation \cite{ZhangKrolikowski:21,HuZhu:20,LengZhu:11,BarbieriCabello:06} are not applicable to waveguides, which will occupy a major area in the future of the field. For similar reasons, we will consider solely collinear phase-matching processes. We will further omit techniques based on the cascading of multiple, distinct nonlinear elements, used to construct e.g. nonlinear interferometers \cite{ChekhovaOu:16}, and techniques relying on integration with other components, like, for instance, the resonators used to create quantum optical microcombs \cite{KuesMorandotti:19} or optical parametric oscillators \cite{CollettGardiner:84,BruchTang:19}. 

We note, that the three-wave-mixing processes discussed in the following can further be influenced by dispersion-engineering \cite{JankowskiFejer:21,WeissPeruzzo:25,WeissPeruzzo(2):25} and pump-shaping techniques \cite{AnsariSilberhorn:18}, which we will not analyze in detail, but which are utilized in some of the schemes presented here.

\bigskip

\begin{figure*}
    \centering
    \includegraphics[width=2\columnwidth]{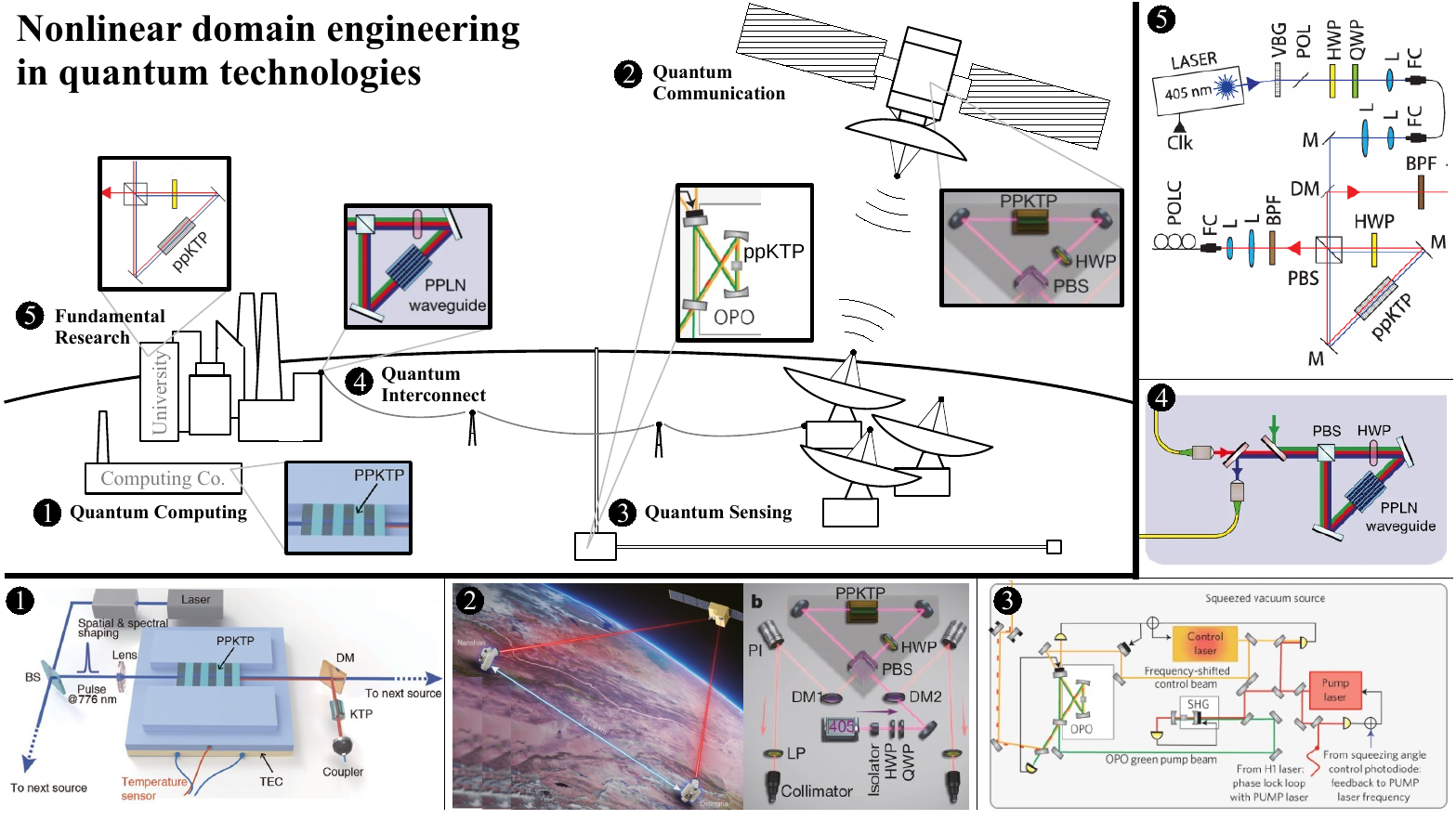}
    \caption{Domain engineered crystals in photonic quantum technologies, including (1) the demonstration of quantum computational advantage (Reproduced with permission from Science 370, 1460 (2020). Copyright 2020 AAAS) (2) satellite-to-ground quantum cryptography (Reproduced with permission from Nature 582, 501 (2020). Copyright 2020 Springer Nature), (3) sensitivity enhancement of LIGO (Reproduced with permission from Nature Photonics 7, 613 (2013). Copyright 2013 Springer Nature), (4) interconnection of disparate quantum memories (T. van Leent \textit{et al.} Nature 607, 69 (2022); licensed under a Creative Commons Attribution (CC BY) license) and (5) significant-loophole-free test of Bell’s theorem (M. Giustina \textit{et al.} PRL 115, 250401 (2015); licensed under a Creative Commons Attribution (CC BY) license). With the exception of (1) all of these experiments still feature strictly periodic design, and do not yet take advantage of the full freedom-of-design available to the domain engineering approach.}
    \label{fig:DEinQT}
\end{figure*}

\noindent
\textbf{Domain engineering in quantum technologies}\\
Domain engineered nonlinear crystals, operated in different regimes and incorporated into different ancillary systems, enable four primary technologies:

\noindent
(\textit{1}) \textit{Single photon generation:} Heralded photons generated by SPDC constitute a highly flexible, bright and easy-to-use single photon source whose probabilistic nature can be overcome using multiplexing techniques \cite{MeyerMigdall:2020}.

\noindent
(\textit{2}) \textit{Entangled photon generation:} Using appropriate setups, SPDC bi-photons are readily generated as maximally entangled states in polarization, time-frequency, time or spatial degrees-of-freedom.

\noindent
(\textit{3}) \textit{Squeezed state generation:} Operated in the high-gain regime, (spontaneous) parametric down-conversion generates squeezed states with squeezing of up to $\sim 15 \mathrm{dB}$.

\noindent
(\textit{4}) \textit{Quantum frequency conversion:} Sum- and difference-frequency generation processes mediate THz-range single photon frequency conversion without undesired changes to other state properties. 

Leveraging these capabilities with the intrinsic promise of room-temperature, no-noise optical implementation and the stability and efficiency promised by integrated circuits, domain-engineered nonlinear crystals present a highly attractive resource, and have found application in milestone experiments across the full breadth of photonic quantum technologies, as illustrated in see Fig. \ref{fig:DEinQT}.

In particular, experiments demonstrating quantum computational advantage, in both discrete \cite{ZhongPan:20} and continuous variable \cite{MadsenLavoie:22} regimes, have relied on (a)periodically poled KTP crystals for state generation. In the field of quantum communication, periodically poled KTP was used to distribute entanglement between satellite and ground stations \cite{YinPan:20}, and periodically poled LN has been employed to demonstrate multi-user quantum secure direct communication \cite{QiChen:21}. Quantum sensing based on squeezed states generated from periodically poled KTP was used to enhance the sensitivity of LIGO \cite{AasiZweizig:13}, interconnect elements based on periodically poled LN waveguides were used to entangle disparate quantum memories \cite{VanLeentWinfurter:22} and milestone experiments investigating fundamental physics utilized KTP for photon generation \cite{GiustinaZeilinger:15}. 

Although already applied successfully across a broad range of applications, the majority of cutting-edge experiments still feature strictly periodically-poled crystals, utilizing little of the available freedom in manipulating the material nonlinearity and designing the associated nonlinear interactions.

In the following we point toward the potential of designed, aperiodical structuring of the material nonlinearity, and show how it may be employed to realize a number of very different objectives, applicable to a broad range of applications.

\bigskip

\begin{figure*}
    \centering
    \includegraphics[width=2\columnwidth]{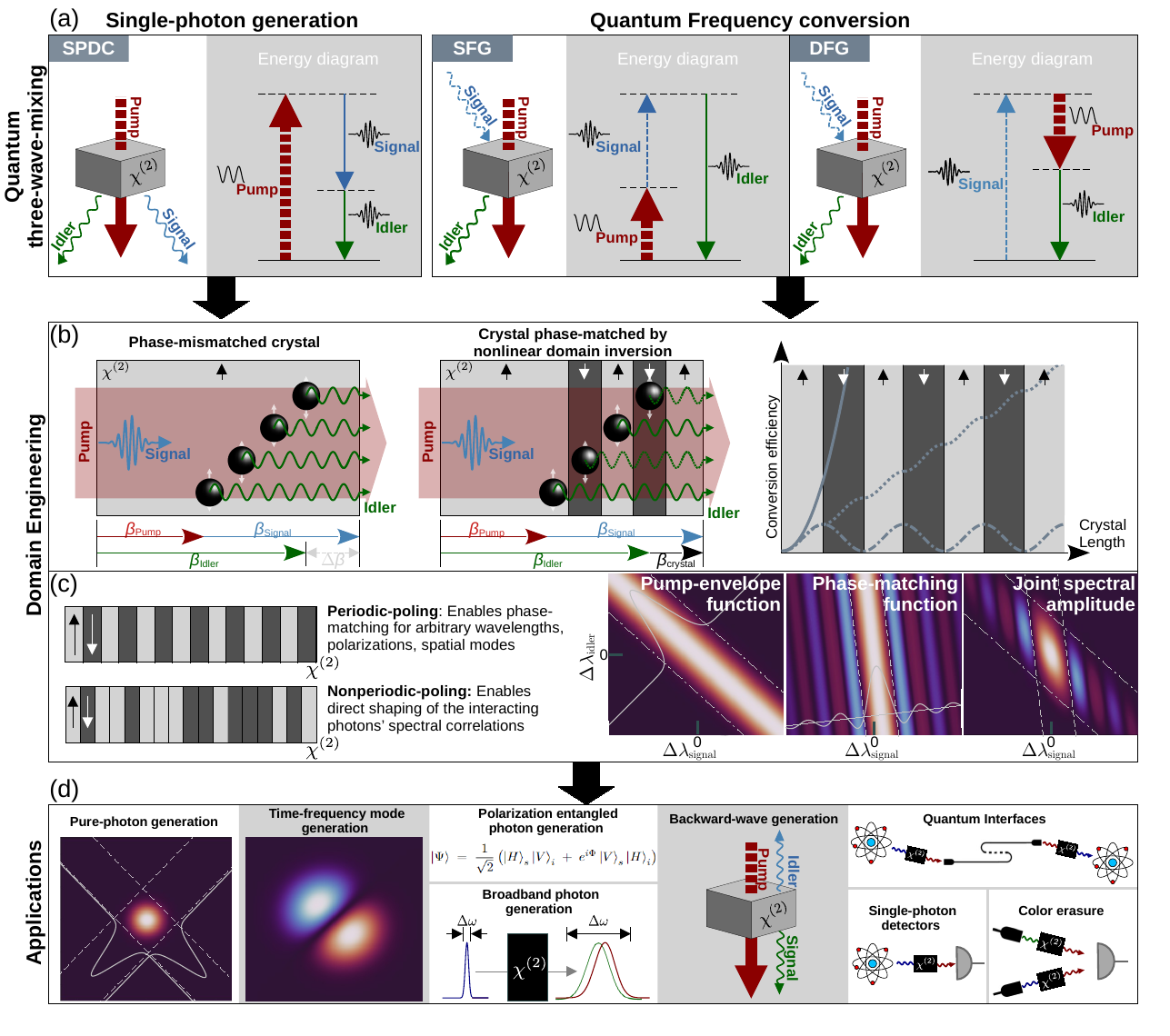}
    \caption{(\textbf{a}) Schematic depiction of quantum regime three-wave-mixing processes: Photons from a strong, classical pump field in a crystal with $\chi^{\mathrm{(2)}}$ nonlinearity can spontaneously decay into single-photon pairs of lower energy, dubbed signal and idler (SPDC). If in addition to the pump field, a single signal-photon is incident, the nonlinearity mediates conversion of the signal-photon and a photon from the pump to a photon at the sum or difference of the respective energies (SFG/DFG). (\textbf{b}) To occur efficiently all processes are required to satisfy conservation of energy and momentum. Momentum conservation, e.g. in terms of waveguide propagation constants $\beta_{k}$, is readily achieved by locally inverting the material nonlinearity to form domains in which the sign of the interaction is reversed, introducing an artificial momentum component in the crystal. For an illustrative, phase-mismatched SFG process (left), light generated by the material dipoles adds incoherently. By inverting the material nonlinearity at appropriate intervals, phase-matching can be achieved artificially (middle). A comparison between a naturally phase-matched process (solid line), an process phase-matched using periodic poling (dotted line), and a phase-mismatched process (dashed-dotted line), are depicted in the plot on the right. (\textbf{c}) Specific arrangement of  nonlinear domains in a (non)periodic fashion allows direct control of the phase-matching function of the process (left). The phase-matching function, together with the pump envelope function, given by the spectral shape of the pump field, fully determines the spectral properties of the interacting signal and idler photons, described by their joint spectral amplitude (right). The pump envelope function can be approximated by a Gaussian corresponding to its spectral bandwidth, and the phase-matching function is given by the Fourier transform of the nonlinearity, corresponding to a sinc-function for a uniformly or strictly-periodically poled crystal. To satisfy energy-conservation, the pump-envelope-function is always orientated along the (anti)diagonal, whereas the angle of the phase-matching function is determined by the relative magnitudes of the group-velocities of the interacting photos, and can be aligned along the diagonal or either of the axes under specific group-velocity-matching conditions. The axes are labeled in terms of detuning from perfect phase-matching. (\textbf{d}) Schematic depictions of the different applications of domain engineering in quantum technologies, as discussed in this paper.}
    \label{fig:ConceptualFigure}
\end{figure*} 

\bigskip

\noindent
\textbf{Three-wave-mixing in the quantum regime}\\
The material response to incident electromagnetic radiation can efficiently be described by a time-varying dipole moment, which, for sufficiently high fields, includes non-linear driving terms, best expressed by expansion of the polarization $P$ into a power series of the interacting electric fields $E$ \cite{Boyd:08}
\begin{equation}
    \begin{aligned}
    \label{eq:PowerSeries}
        P_{j}(t) \; = \;  & \epsilon_{0} \biggl[ \overbrace{ \sum_{k} \chi_{jk}^{\mathrm{(1)}} E_{k} (t)}^{\text{linear optics}} \; + \; \overbrace{ \sum_{kl} \chi_{jkl}^{\mathrm{(2)}} E_{k} (t) E_{l} (t) }^{\text{three-wave-mixing processes}} \\
        & + \; \underbrace{ \sum_{klm} \chi_{jklm}^{\mathrm{(3)}} E_{k} (t) E_{l} (t) E_{m} (t) }_{\text{four-wave-mixing processes}} \; + \; ... \; \biggr] .
    \end{aligned}
\end{equation}
Here, the first term describes the harmonic interaction responsible for diffraction, whereas the higher-order terms correspond to the material's nonlinear response, and act as sources of new components of the electromagnetic field. The three-wave mixing processes discussed here are mediated by the $\chi^{\mathrm{(2)}}$-nonlinearity, which features nonlinear interaction notably stronger than higher-order terms.

Three-wave-mixing features a number of different possible interactions, but, within the scope of this paper, we are interested only in the processes of SPDC, generating single photon pairs, and quantum frequency conversion (QFC), in which a single-photon input is transferred to a new frequency-band. The corresponding Hamiltonians are readily derived from (\ref{eq:PowerSeries}) using Poynting's theorem and appropriate representation of the electric fields \cite{GriceWalmsley:97,RubinSergienko:94,ODonnelURen:09,ChristSilberhorn:13}:
\begin{widetext}
    \begin{equation}
    \begin{aligned}
        \hat{H}_{\mathrm{QFC/\gray{SPDC}}}(t) \; = \; C \; & \highlightblue{ \iint \mathrm{d}x \mathrm{d}y \; \chi^{\mathrm{(2)}}  f_{p} (x,y) f_{s}^{\gray{*}} (x,y) f_{i}^{*} (x,y) } \; \highlightgreen{ \iiint \mathrm{d}\omega_{p} \mathrm{d}\omega_{s} \mathrm{d}\omega_{i} \; \mathrm{e}^{-i \Delta \omega t} } \\ 
        & \overbrace{ \underbrace{ \highlightyellow{ \int_{} \mathrm{d}z \; s_{\mathrm{nl}}(z) \mathrm{e}^{i\Delta \beta (\omega_{p}, \omega_{s}, \omega_{i}) z} } }_{\text{Phase-matching function} \; \Phi} \; \alpha(\omega_{p}) }^{\text{Joint spectral amplitude}}\hat{a}_{s}^{\gray{\dagger}} (\omega_{s}) \hat{a}_{i}^{\dagger} (\omega_{i}) \; \; + \; \; H.c. ,
    \end{aligned}
    \end{equation}
\end{widetext}
wherein the single-photon signal and idler fields are quantized, while the strong pump with spectral distribution $\alpha(\omega_{p})$ is treated classically; all constants are incorporated into the factor $C$. The representations of SPDC and QFC differ only in the form the signal photon is included, using creation $\hat{a}_{k}^{\dagger}$ and annihilation $\hat{a}_{k}$ operators at frequencies $\omega_{k}$ respectively. We note, that the Hamiltonian of SPDC formally corresponds to a two-mode squeezing operator.

The term highlighted in blue describes the nonlinear overlap of the interacting fields $f_{k}(x,y)$. This represents a key parameter related directly to the process efficiency, and requires, in the case of waveguide implementation, an appropriate selection of the respective discrete waveguide spatial modes.

The remaining integrals describe the so-called 'phase-matching' of the interaction. The expression highlighted in green, after appropriate time integration, enforces energy conservation between the photons, requiring the frequency mismatch $\Delta \omega$ to be zero. This can be satisfied by different combinations of the interacting pump, signal and idler photons, leading to different interactions, distinguishing SPDC and QFC processes (see. Fig. \ref{fig:ConceptualFigure}a).

The term highlighted in yellow constitutes the conservation of momentum, here represented by the mismatch $\Delta \beta$ of waveguide propagation constants. The different three-wave-mixing processes occur optimally only if this mismatch is equal to zero. This can generally only be satisfied for one process at a time, which we will refer to in the following as the phase-matched process. At what frequencies phase-matching is achieved, for which photons, in terms of polarization or spatial modes, it occurs, and how the conversion-probability is 'shaped'  in frequency space, can be controlled directly by structuring the material nonlinearity. In fact, the shape of the so-called phase-matching function $\Phi$, which is, together with the so-called pump-envelope function $\alpha(\omega_{p})$, responsible for the spectral shape of the interacting photons, is given directly by the Fourier transform of the nonlinearity, here expressed via its dependence on the propagation direction $s_{\mathrm{nl}}(z)$.

Specifically, a periodic inversion of the nonlinearity allows to introduce an artificial wave-vector into the expression of the momentum mismatch $\Delta \beta$ (see Fig. \ref{fig:ConceptualFigure}b), enabling phase-matching of photons at nearly arbitrary frequencies, polarization and spatial modes \cite{HumFejer:07}. Custom, non-periodic structuring of the material can further be used to directly shape the spectral properties of the generated/converted photons \cite{AnsariSilberhorn:18}. This can be illustrated by calculating the state generated from SPDC using appropriate time evolution, assuming weak pumping, and performing Taylor expansion to first order, to arrive at the two-photon state \cite{ChristSilberhorn:13,Branczyk:10}
\begin{equation}
    \label{eq:SPDCState}
    \ket{\Psi} \; = \; \iint \mathrm{d}\omega_{s} \mathrm{d}\omega_{i} \; \underbrace{ \Phi (\omega_{s},\omega_{i}) \alpha (\omega_{s} + \omega_{i}) }_{\text{Joint spectral amplitude}} \hat{a}_{s}^{\dagger} (\omega_{s}) \; \hat{a}_{i}^{\dagger} (\omega_{i}) \; \ket{00}.
\end{equation}
From (\ref{eq:SPDCState}) it can be seen directly, that the spectral properties of the generated photon pair are determined solely by the phase-matching function $\Phi$ and the pump envelope function $\alpha(\omega_{s} + \omega_{i})$. A conceptual depiction of how both functions shape the spectral properties of the generated bi-photon, a feature to which we will in the following referred to as the joint-spectral-amplitude (JSA), is shown in Fig. \ref{fig:ConceptualFigure}c).

The JSA can generally not be separated with respect to the individual photons, indicating intrinsic spectral correlations. The internal structure of these correlations can be revealed by performing Schmidt-decomposition of the JSA and rewriting the state using broadband-mode creation operators $\hat{A}_{k}^{\dagger}$/$\hat{B}_{k}^{\dagger}$ instead of the monochromatic ones employed in (\ref{eq:SPDCState}) \cite{URenRaymer:06}:
\begin{equation}
    \label{eq:JSA}
    \mathrm{JSA} \: {\overset{\begin{subarray}{l} \text{\; \; Schmidt}\\ \text{decomposition} \end{subarray}}{=\joinrel=\joinrel=\joinrel=}} \: \sum_{k} \lambda_{k} u_{k}(\omega_{s}) v_{k}(\omega_{i}) 
\end{equation}
\begin{equation}
    \label{eq:BroadbandModePicture}
    \ket{\Psi} \; = \; \sum_{k} \lambda_{k} \underbrace{ \int \mathrm{d}\omega_{s}  u_{k}(\omega_{s}) \hat{a}_{s}^{\dagger} }_{\hat{A}_{k}^{\dagger}} \underbrace{ \int \mathrm{d}\omega_{i} v_{k}(\omega_{i}) \hat{a}_{i}^{\dagger} }_{\hat{B}_{k}^{\dagger}} \ket{00}
\end{equation}
Now, the state in (\ref{eq:BroadbandModePicture}) reveals the SPDC bi-photon state as a superposition of broadband wavepackets, represented by the Schmidt functions $u_{k}(\omega_{s}), v_{k}(\omega_{i})$. The amount of entanglement between the signal and idler photons can be inferred directly from the Schmidt coefficients $\lambda_{k}$, and represents another engineerable quantity, which is desirable in schemes like entanglement distribution, and unwanted in applications based on heralding.

In the high gain regime, utilizing a strong pump, squeezer characteristics begin to dominate over photon-pair characteristics in the SPDC output state, resulting in the generation of squeezed states \cite{WuWu:86}. Accordingly, the first order approximation used to arrive at (\ref{eq:SPDCState}) is no longer accurate, and states corresponding to multiple pair generation need to be included. Furthermore, time ordering effects, which were previously neglected to allow for Taylor series expansion, become relevant \cite{Branczyk:10,ChristSilberhorn:13,QuesadaSipe:22}. Squeezed states themselves are best described using an extended theoretical description \cite{LoudonKnight:87,QuesadaSipe:22,JankowskiFejer:24,BelloPeer:21}.

In the following we will consider SPDC predominantly as a method of generating single-photon pairs, a regime in which high-gain effects can be neglected. QFC, however, is generally operated in the high gain limit, as the power of the pump field directly relates to the conversion efficiency. Nonetheless, the state after conversion can effectively be described by (\ref{eq:SPDCState}) upon replacing the signal creation with an annihilation operator and including the corresponding input photon in the initial state. QFC High-gain effects impact conversion efficiencies and spectral structures, and will result in moving from single-mode conversion toward conversion into multiple distinct modes \cite{ChristSilberhorn:13,QuesadaSipe:14}.

\bigskip

\noindent
%\begin{center}
\textbf{Applications of nonlinear domain engineering}\\
%\end{center}
The domain engineering approach was first developed theoretically well over fifty years ago \cite{ArmstrongPershan:62,FrankenWard:63}, but, due to the difficulty of the associated fabrication, did not see widespread adoption until the late 80s. Since then, in parallel with steadily evolving fabrication capabilities, the technique has moved from strictly periodic structures to involve aperiodic designs tailored toward their respective application. First developed for classical frequency conversion \cite{FengMing:90,ZhuMing:97,ChouBrener:99,TehranchiKashyap:08}, and used to generate optical frequency combs to this date \cite{WuDiddams:24}, this approach was soon adopted to develop sophisticated designs suitable for applications in quantum technologies. These designs can roughly be separated into three categories:

(\textit{1}) Approaches manipulating bi-photon entanglement in the time-frequency degree-of-freedom by directly shaping the phase-matching function in Fourier-space. 

(\textit{2}) Approaches aiming to satisfy multiple phase-matching conditions simultaneously via, for instance, multi-periodic structures or chirped gratings.

(\textit{3}) Approaches relying on specifically designed periodic gratings to generate photons in particular polarization, frequency and spatial modes, or to satisfy exotic conditions like backward-wave generation.

In the following, we will one-by-one address the applications to which each of these approaches has been utilized to date. We will focus first on domain engineering based on SPDC, and move from applications based on well-developed approaches towards efforts with the technique still in its infancy.

\bigskip

\begin{figure*}
    \centering
    \includegraphics[width=2\columnwidth]{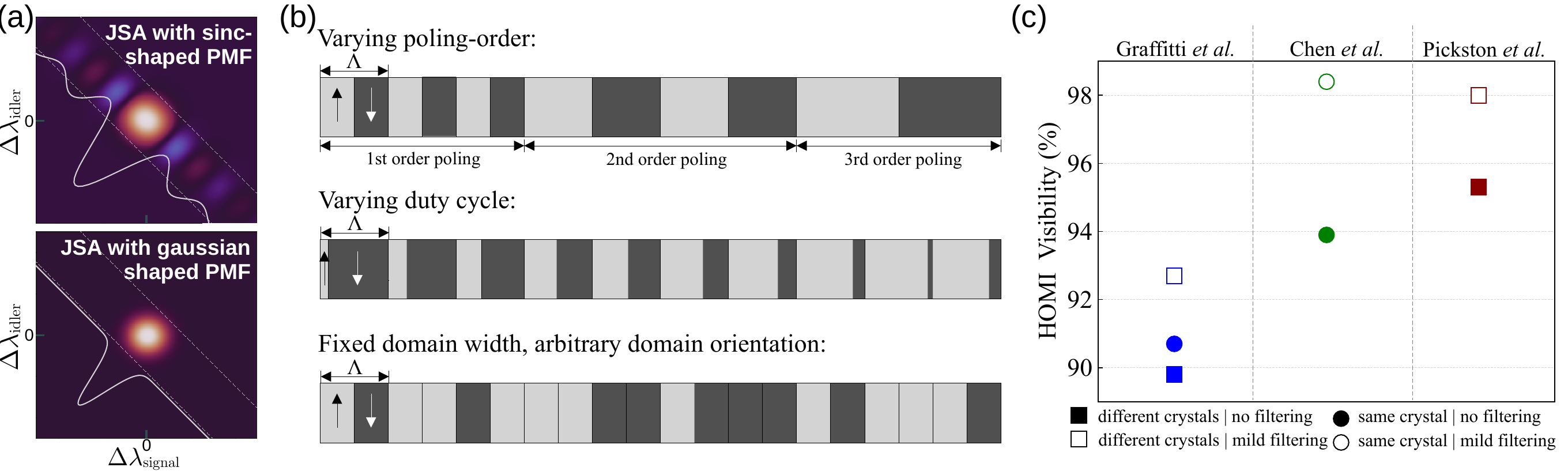}
    \caption{\textbf{Pure photon generation} (\textbf{a}) Joint-spectral-amplitudes of down-converted signal and idler photons for a sinc-shaped phase-matching function, which is the result of an un- or periodically-poled crystal (top), and the ideal Gaussian phase-matching function (bottom). The side-lobes of the sinc-function result in spectral correlations, reducing single-photon purity after heralding. The depicted joint-spectral-amplitudes are obtained for a Gaussian pump at symmetric group-velocity-matching conditions, which represent additional conditions for separability. (\textbf{b}) Schematic illustration of the available approaches for generating Gaussian phase-matching functions. (\textbf{c}) Comparison between the highest single-photon purities achieved to date, including Graffitti \textit{et al.} \cite{GraffittiFedrizzi:18}, Chen \textit{et al.} \cite{ChenWong:19} and Pickston \textit{et al.} \cite{PickstonFedrizzi:21}. The purity of the photons generated from the crystal in \cite{PickstonFedrizzi:21} is almost 15\% higher than that of an unfiltered, periodically poled crystal. The heralded single-photon purity is given directly by the visibility of the heralded (four-fold coincidence) Hong-Ou-Mandel interference \cite{Branczyk:17}; other methods can lead to an overestimation of the purity. Both heralded interference of successive photons from the same source and photons from different sources are possible, and can be up-scaled in temporal and spatial multiplexing architectures respectively. It is possible to further increase the single-photon purity by employing mild spectral filtering, affecting the heralding efficiency by as little as $<1\%$ \cite{PickstonFedrizzi:21}.}
    \label{fig:PurePhotonGeneration}
\end{figure*}

\noindent
\textbf{Pure photon generation}\\
A large number of photonic quantum technologies, among them measurement-based quantum computing \cite{RaussendorfBriegel:03,WaltherZeilinger:05}, boson-sampling \cite{BroomeWhite:13,MeerPinkse:20} and photonic quantum repeaters \cite{AzumaLo:15}, are based on the quantum-mechanics of two-photon interference. This interference occurs optimally only if the involved photons exist in spectrally pure states \cite{Branczyk:17}. Sub-optimal interference can be compensated only by increasing circuit complexity and the number of required photon sources and detectors \cite{LiBenjamin:10,MeyerSilberhorn:17}. 

As single-photon sources based on SPDC generally rely on heralding and multiplexing techniques to overcome their probabilistic nature, the spectral correlations of the down-converted photon pair impose an upper limit on the purity of a heralded, single photon. This can be seen directly from (\ref{eq:BroadbandModePicture}) by calculating the state after detection of the idler, upon which the signal is projected into a statistical mixture of its broadband frequency modes. While signal/idler correlations can in principle be addressed via spectral filtering, such techniques are directly associated with a decrease in heralding efficiency and source brightness \cite{URenRaymer:06,Branczyk:10}. Furthermore, applications directly utilizing the generated bi-photon states, for interference or as polarization-entangled qubits, prefer a separable JSA to avoid contamination during interference \cite{GraffittiBranczyk:18}. Accordingly, spectral separability represents an often highly desirable quality, and should be incorporated into the source design of many applications.

Separability is achieved only if the pump-envelope function, given by the spectral properties of the pump pulse, and the phase-matching function are designed correctly. The latter, in particular, is required to be a Gaussian \cite{QuesadaBranczyk:18}, which can be achieved by careful manipulation of the sign of the material nonlinearity.

This was first demonstrated in 2011 \cite{BranczykWhite:11} utilizing different order periodic-poling, followed by alternative, improved approaches based on variation of the poling duty-cycle \cite{DixonWong:13} and arbitrary arrangement of domains of fixed width \cite{DossevaBranczyk:16,TambascoMitchell:16}. For both of the latter approaches high single-photon purity inferred from heralded interference has been demonstrated experimentally \cite{GraffittiFedrizzi:18,ChenWong:19,PickstonFedrizzi:21} achieving up to $>98\%$ purity without filtering.

There further exist a number of theoretical works, extending the methods \cite{GraffittiFedrizzi:17}, presenting algorithms to optimize the poling designs \cite{CuiZhang:19,CaiJin:22} and considering fabrication errors \cite{GraffittiBranczyk:18}. A similar domain engineering approach for pure photon generation was employed in a milestone experiment demonstrating photonic quantum computational advantage \cite{ZhongPan:20}.

Fig. \ref{fig:PurePhotonGeneration} presents a schematic depiction of the different approaches, together with a summary of the highest achieved single-photon purities to date.

So far, all experiments demonstrating domain-engineered pure photon generation have utilized bulk KTP crystals. Adapting the different techniques to other material platforms or waveguide implementation will require reevaluation of the underlying phase-matching conditions. In particular group-velocity-matching conditions \cite{URenRaymer:06}, a key requirement for the separability of (\ref{eq:JSA}), is given directly by the material dispersion, which can differ markedly between different platforms, and may limit the wavelength availability of pure photons generated in this fashion \cite{ZhuJin:23,CaiJin:22,MccrackenFedrizzi:18}.

To end this section, we would like to shortly discuss the different approaches of manipulating the phase-matching function, depicted in Fig. \ref{fig:PurePhotonGeneration}b), as well as the algorithmic approaches developed to design them: The approaches based on varying poling order \cite{BranczykWhite:11} and duty-cycle variation \cite{DixonWong:13}, while intuitive and well understood from applications in classical nonlinear optics, rely on global optimization algorithms which can be cumbersome and not always optimal. The approach featuring arbitrary domain orientation \cite{DossevaBranczyk:16} has been developed to operate deterministically \cite{TambascoMitchell:16} using domains of arbitrary size \cite{GraffittiFedrizzi:17}, and, as a result, can be used to approximate the target function up to the physical limits of the problem. As result, recent works almost exclusively rely on algorithms akin to the one described in \cite{GraffittiFedrizzi:17}.

\bigskip

\begin{figure*}
    \centering
    \includegraphics[width=2\columnwidth]{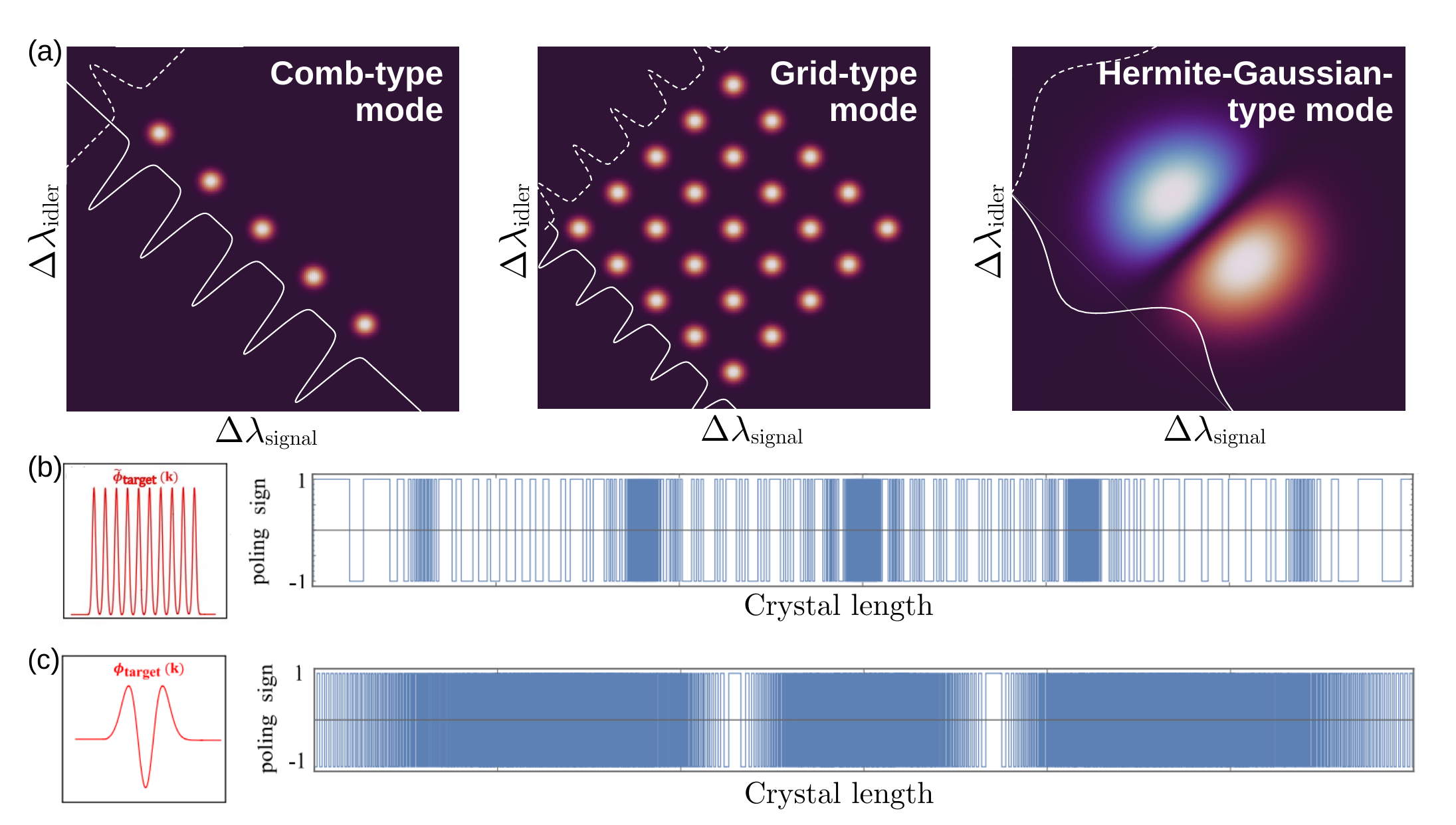}
    \caption{\textbf{Time-frequency mode generation} (\textbf{a}) Schematic depiction of different TFMs. At symmetric group-velocity-matching, the spectral shape of the signal and idler photons is given by the product of the phase-matching function (orientated along the diagonal) and the pump envelope function (orientated along the anti-diagonal), as indicated by the solid and dashed curves respectively. Comb- and grid-like states can be generated from a multi-peaked phase-matching function. Hermite-gaussian-type TFMs require a phase-matching function shaped like the corresponding Hermite-gaussian. If the phase information is considered, these modes are field overlapping, but orthogonal. Exemplary designs of poling periods, corresponding to a multi-peaked (\textbf{b}) and 2nd order Hermite-gaussian (\textbf{c}) phase-matching function (Reproduced with permission from JOSA B 40, A9 (2023). Copyright 2023 Optical Society of America). Here, the approach with arbitrary domain orientation (see. Fig. \ref{fig:PurePhotonGeneration}b) was employed.}
    \label{fig:TFMGeneration}
\end{figure*}

\noindent
\textbf{Time-frequency mode generation}\\
Information may be encoded into photons via any, or any combination, of their degrees of freedom. These include spatial modes, realized either via distinct paths or transverse mode profiles encoded in orbital angular momentum, polarization, or their structure in the time-frequency domain. Encoding in the time-frequency degree-of-freedom promises an, in principle, infinite-dimension Hilbert space as well as compatibility with fiber-networks \cite{AnsariSilberhorn:18(2)}, both features currently unavailable to polarization and spatial encoding schemes, respectively.

High-dimensionality is especially attractive for quantum communication technologies, where it may be used to increase packing density via time-frequency modes (TFM) multiplexing, or to directly encode information into superpositions of different TFM \cite{BrechtRaymer15}, enabling genuine high-dimensional quantum communication to increase channel capacity and security \cite{BechmannTittel:00,CerfGisin:02}. Moreover, the set of single-qubit gates required for linear optical quantum computing \cite{KnillMilburn:01,KokMilburn:07} may be implemented in time-frequency space using so-called quantum-pulse-gates \cite{EcksteinSilberhorn:11}, and graph-states with tailored entanglement structure for cluster-state quantum computing may be generated via fusion of Bell pairs \cite{BrechtRaymer15,FabreMilman:20}.

TFMs can be encoded in temporally and spectrally distinct time- \cite{BrendelZbinden:99} or frequency-bins \cite{RamelowZeilinger:09,MalteseDucci:20,AppasDucci:21}, frequency and time overlapping, but field orthogonal pulse modes \cite{LawEberly:00}, or combinations of the same. Many of these states can be generated directly at the source, by engineering the spectral-temporal shape of either the pump field, or the phase-matching function of the down-conversion process. The latter can be achieved by appropriate design of the material non-linearity, in analogy to the approach aiming to generate pure photons described above, but with a modified target function. This represents a stable and controlled alternative to classical methods of generating TFMs \cite{PeerSilberberg:05,AverchenkoLeuchs:17,Matsuda:16,FrancesconiDucci:20}.

The approach was first demonstrated by Graffitti \textit{et al.} \cite{GraffittiFedrizzi:20(1)}, generating Bell states encoded in a pulsed TFM by engineering the phase-matching function to match a first-order Hermite-Gaussian. The scheme was later extended to demonstrate hyperentanglement between TFM and polarization/orbital-angular-momentum degrees-of-freedom \cite{GraffittiFedrizzi:20(2)}. Phase-matching functions domain-engineered to a comb-like structure were further utilized to generate frequency-bin/polarization hyperentanglement \cite{MorrisonFedrizzi:22}, and, using temperature control to shift the generated photon spectra, hyperentanglement between pulse-modes and frequency bins (both in the time-energy degree-of-freedom) was achieved \cite{ChirianoFedrizzi:23}. Recently, TFM generation utilizing both engineering of the phase-matching function and shaping of the pump pulse has been demonstrated \cite{ShukhinEisenberg:23}. 

There further exist works presenting theory and simulations, proposing to extend the approach to the generation of bright-squeezed-vacuum multi-mode grid-type states \cite{DragoBranzcyk:22,HurvitzArie:23}, and the mid-infrared region \cite{ZhuJin:23}.

Experimental demonstrations employ both the domain-engineering approach relying on arbitrary domain orientation and on varying duty cycles (see Fig. \ref{fig:PurePhotonGeneration}b). So far, only bulk KTP crystals phase-matched under symmetric group-velocity-matching have been utilized. We present exemplary TFMs together with examples of the poling structures used to implement them in Fig. \ref{fig:TFMGeneration}.
 
\bigskip

\begin{figure}
    \centering
    \includegraphics[width=1\columnwidth]{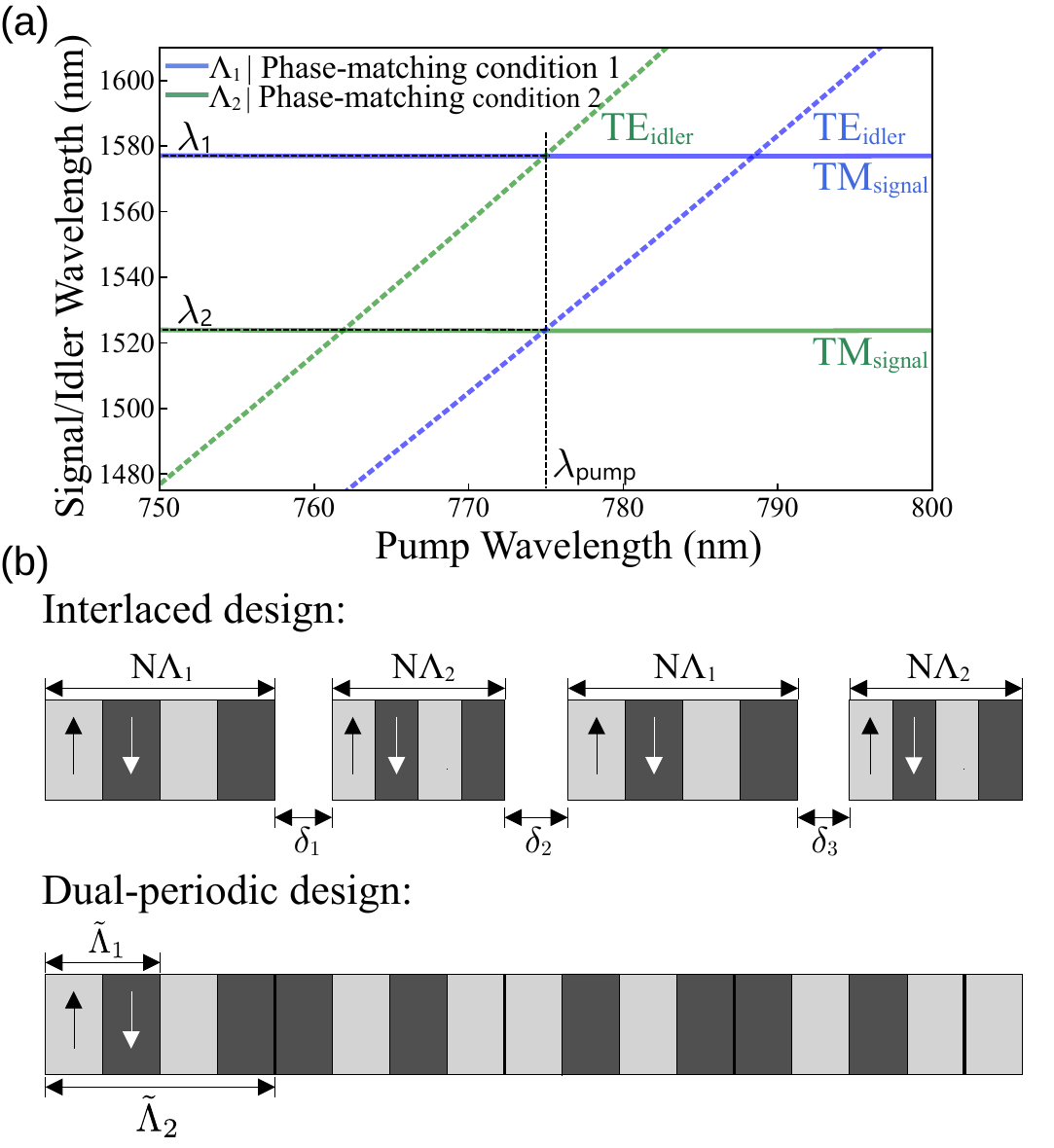}
    \caption{\textbf{Polarization-entangled photon generation} (\textbf{a}) Schematic depiction of two $\mathrm{TM_{pump} \longrightarrow TM_{signal} \; + \; TE_{idler}}$ down-conversion processes suitable for the generation of polarization-entangled photons ($\mathrm{TM}/$$\mathrm{TE}$ describe orthogonal polarization modes). Pumping the corresponding crystal at $\lambda_{p}$ will result in generation of one of the two photon pairs $\mathrm{TM_{\lambda_{1}}}, \mathrm{TE_{\lambda_{2}}}$ or $\mathrm{TE_{\lambda_{1}}}, \mathrm{TM_{\lambda_{2}}}$; equivalent to the $\Psi^{+/-}$ Bell-state. (\textbf{b}) Poling structures capable of phase-matching a crystal to two processes simultaneously, utilizing an interlaced design (top) and a dual-periodic design (bottom). The interlaced design directly enables the down-conversion processes phase-matched by the poling-periods $\Lambda_{1,2}$. The spacings $\delta$ are to be chosen so that the distance between sections of the same period is a multiple of this period. The dual-periodic design enables processes phase-matched by the reciprocal vector $G_{mn} = \frac{2 \pi m}{\tilde{\Lambda}_{1}} + \frac{2 \pi n}{\tilde{\Lambda}_{2}}$, wherein $m,n$ represent the (possibly negative) phase-matching order.}
    \label{fig:PolarizationEntangledPhotonGeneration}
\end{figure} 

\noindent
\textbf{Polarization-entangled photon generation}\\
The distribution of entanglement between distant parties constitutes a key component in efforts toward quantum cryptography systems \cite{GisinZBinden:02} and large-scale quantum networks \cite{Kimble08}. To date, this has most readily been achieved utilizing polarization-entangled photon pairs \cite{WengerowskyUrsin:18,YinPan:17,YinPan:20}.

Polarization-entangled photon sources are generally based on either bidirectional pumping of a crystal phase-matched to a single down-conversion process, or separate pumping of two such crystals \cite{KimWong:06,SansoniSilberhorn:17,AtzeniOsellame:18}. Both approaches require either bulk-optics interferrometers, complex integrated components, or generate the desired states only probabilistically.

The generation of wavelength non-degenerate polarization-entangled photons can alternatively be achieved utilizing a crystal simultaneously phase-matched to two different phase-matching conditions, as illustrated in Fig. \ref{fig:PolarizationEntangledPhotonGeneration}.

This was first proposed in 2009 \cite{SuharaFujimura:09,ThyagarajanTanzilli:09} and later demonstrated, in principle, for waveguides structured with interlaced sections of different poling periods \cite{HerrmannSilberhorn:13}. The approach was since developed to involve designs based on dual-periodic poling structures, for integrated waveguides \cite{SunNing:19} and bulk crystals \cite{KuoNam:20}, achieving a source brightness of up to $1.22 \times 10^{7} \mathrm{ pairs \; mW^{-1} nm^{-1} s^{-1}}$ and interference visibility of $94 \%$. To date, all demonstrations utilized LN.

The photons generated by these schemes are fully suitable for operations in which spectral indistinguishability of the photons is not required. In particular, applications in which one photon is designated for long-range transmission through optical fibers at telecommunication wavelengths, and the other is used for local, largely wavelength-independent operations, or for storage in a memory, where the wavelength is required to match e.g. an atomic transition.

Extending this approach to wavelength-degenerate applications requires an alternate method of separating the generated photons, which may be achieved by phase-matching two counter-propagating down-conversion processes \cite{GongZhu:11}.

\bigskip

\begin{figure}
    \centering
    \includegraphics[width=1\columnwidth]{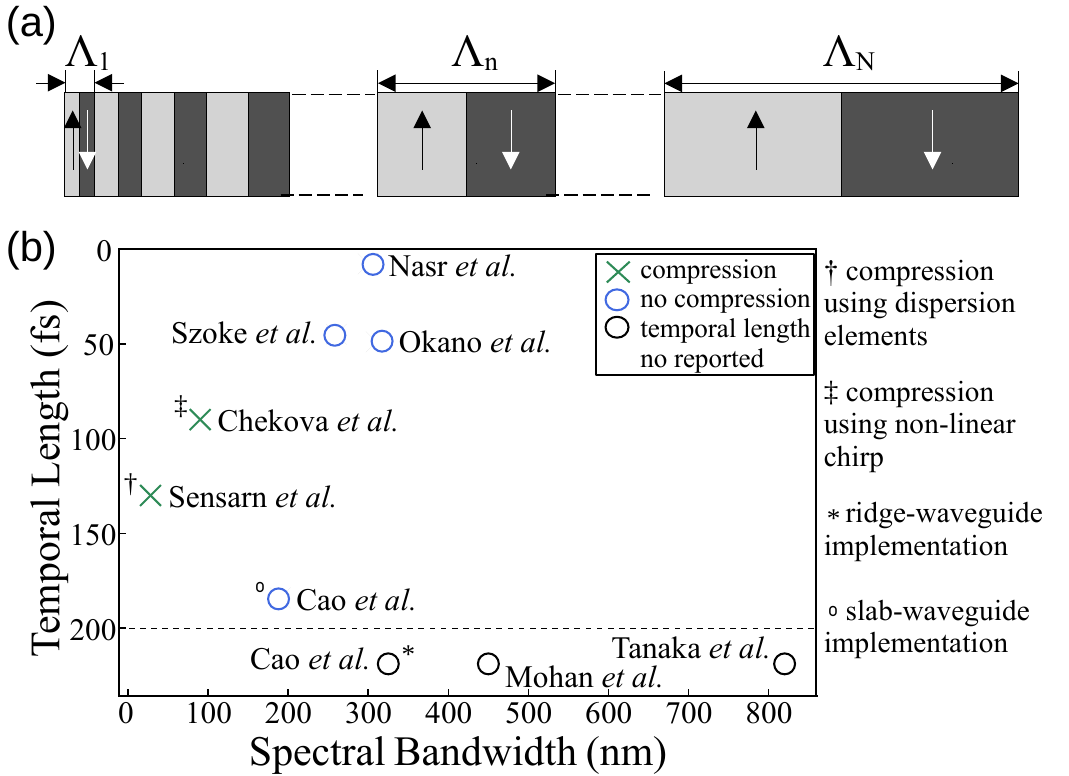}
    \caption{\textbf{Broadband photon generation} (\textbf{a}) Schematic depiction of a chirped poling structure. Each poling period phase-matches a different wavelength, resulting in broadband photon generation. For the linearly chirped structure depicted here, longer wavelengths are generated earlier in the crystal, acquiring an additional phase with respect to shorter wavelengths, resulting in chirped (bi-)photons. While such photons show narrow Hong-Ou-Mandel interference (HOMI) dips, the chirp needs to be removed to compress the photons' wavepacket to, ideally, the transform limit. (\textbf{b}) Comparison of the different broadband photons generated in chirped crystals to date, including Nasr \textit{et al.} \cite{NasrFeyer:08}, Szoke \textit{et al.} \cite{SzokeCushing:21}, Okano \textit{et al.} \cite{OkanoTakeuchi:15}, Chekova \textit{et al.} \cite{ChekhovaPrudkovskii:18}, Sensarm \textit{et al.} \cite{SensarnHarris:10}, Cao \textit{et al.} \cite{CaoTakeuchi:21,CaoTakeuchi:23}, Tanaka \textit{et al.} \cite{TanakaTakeuchi:12} and Mohan \textit{et al.} \cite{MohanTeich:09}. Experiments without compression of the generated photons (blue) determine the temporal correlations using HOMI, which does reflect the true width of the wavepacket, which remains broad. Source brightnesses reach up to $1.2 \cdot 10^{6} \mathrm{pairs \; \mu W^{-1} s^{-1}}$ \cite{CaoTakeuchi:21}. The non-linearly chirped crystal in \cite{ChekhovaPrudkovskii:18} does not produce an approximate flattop spectrum, with the reported bandwidth corresponding to a base-to-base measurement, all other bandwidths are taken at full-width-half-maximum. Unless specified otherwise, all experiments were performed using bulk crystals structured with a linear chirp.}
    \label{fig:BroadbandGeneration}
\end{figure} 

\noindent
\textbf{Broadband photon generation}\\
The spectral and temporal properties of photons generated by SPDC are, due to time-frequency duality, inherently linked. For optimal, transform-limited photons, whose time-bandwidth product is at a minimum, the two-photon correlation time scales inversely with their spectral bandwidth. This, in principle, allows to engineer ultra-short (bi-)photons, coincident within a single optical cycle, by maximizing their spectral bandwidth.  

Such ultra-short correlation times represent a key metric for quantum optical coherence tomography \cite{NasrTeich:03}, clock-synchronization methods \cite{ValenciaShih:04}, or techniques like entangled-photon spectroscopy \cite{SalehTeich:98} and -lithography \cite{BotoDowling:00}, relying on two-photon absorption \cite{DayanSilberberg:04}. Large spectral bandwidths further represent a key parameter for squeezed states generated in the high-gain limit of SPDC.

While there exist different approaches for broadband photon generation, generally relying on either constraining group-velocity-matching conditions or bulk-crystals, the highest bandwidths have been achieved using chirped structuring of the material nonlinearity, effectively phase-matching a broad range of different frequencies, as depicted in Fig. \ref{fig:BroadbandGeneration}. This was first proposed to enhance the axial resolution of quantum optical coherence tomography \cite{CarrascoTeich:04}. Subsequently, the approach was demonstrated in principle \cite{NasrFeyer:08,NasrTeich:08} and application \cite{MohanTeich:09,OkanoTakeuchi:15}, achieving resolution and dispersion tolerance superior to its classical counterpart.

While these demonstrations already yield superior resolution in Hong-Ou-Mandel type interference experiments, they are still affected by a frequency-dependent phase (chirp), resulting from the frequency-dependent point of generation within the crystal. To address this, subsequent compression of the photons' correlation time by removing the frequency chirp was proposed \cite{Harris:07,BridaShumilkina:09} and demonstrated using dispersive optical elements \cite{SensarnHarris:10} and a non-linearly chirped crystal design \cite{ChekhovaPrudkovskii:18}, to achieve, ideally, a transform-limited state.

There further exist theoretical models for the broadband generation of squeezed light \cite{HoroshkoKolobov:13,HoroshkoKolobov:17} and demonstrations of broadband generation using non-collinear emission \cite{TanakaTakeuchi:12} and high-power pumping \cite{SzokeCushing:21}. Waveguide implementations of broadband generation utilizing chirped crystals have recently been demonstrated as well \cite{CaoTakeuchi:21,CaoTakeuchi:23}.

Deviations from a strictly linear chirp, like the one reported in \cite{ChekhovaPrudkovskii:18}, have great potential for further optimization of the technique. Hardware limitations, like broadband, high-efficiency detectors, and material transparency and absorption currently represent the main obstacles of the technique.

\bigskip

\begin{figure}
    \centering
    \includegraphics[width=1\columnwidth]{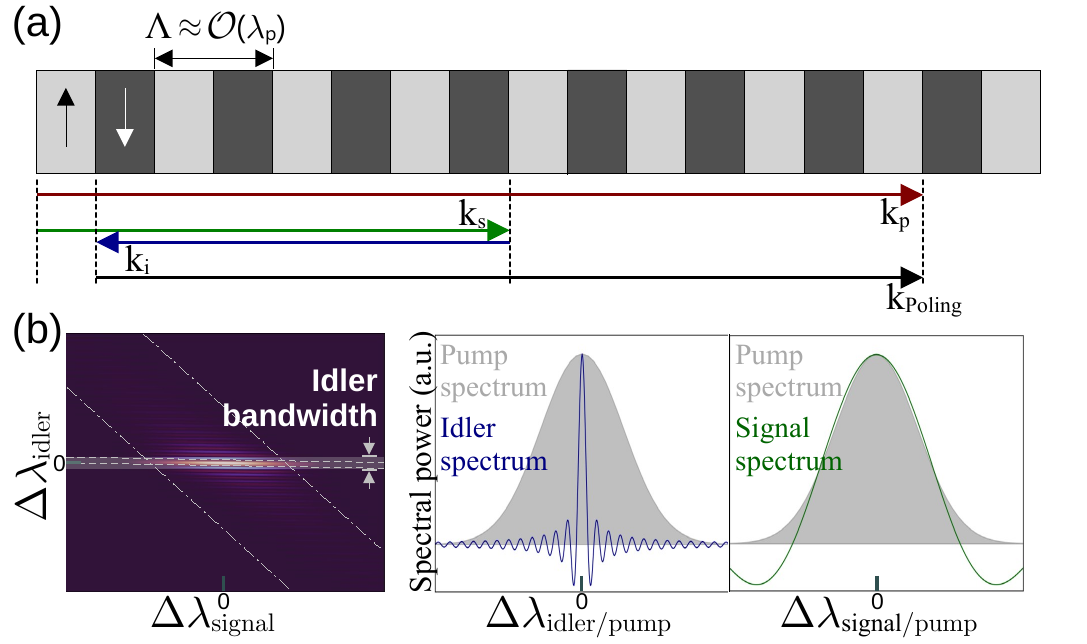}
    \caption{\textbf{Backward-wave SPDC} (\textbf{a}) Sketch of the ultra-short period periodic-poling pattern required for a backward-wave SPDC process, together with the corresponding phase-matching condition: The signal photon is co- and the idler counter-propagating with the pump, leading to a phase-mismatch on the order of the pump wavelength, which has to be compensated using domain engineering. (\textbf{b}) Schematic representation of the JSA of a backward-wave SPDC process, resulting in a highly separable state similar to that achieved using asymmetric group-velocity matching in the co-propagating case (left). Schematic comparison of the bandwidths of the generated idler and signal photons, showing much-reduced bandwidth and spectral distribution following that of the pump, respectively (right).}
    \label{fig:BackwardWaveSPDC}
\end{figure} 

\noindent
\textbf{Backward-wave SPDC}\\
A backward-wave SPDC process generates signal and idler photons propagating in opposite directions. This can be realized using a strictly periodic nonlinear grating with a period on the order of the pump wavelength. While the fabrication of such gratings is still nontrivial, the corresponding phase-matching conditions can alternatively only be fulfilled utilizing huge birefringence and/or a non-copropagating pump.

The counter-propagating geometry gives rise to distinct characteristics, uniquely suited for quantum applications. Most notably, the backward-propagating photon possesses a drastically narrowed spectral bandwidth and a markedly increased temporal length. This is because the time the photon exits the crystal can no longer be inferred from the pump pulse, to within the limit of group-velocity-dispersion, due to the uncertainty about where in the crystal the down-conversion process took place. Narrow photon bandwidths are essential for interfacing with solid-state memories \cite{LvovskyTittel:09,HeshamiSussman:16}, which represent essential components for quantum repeaters \cite{DuanZoller:01,SangouardGisin:11} and linear optics based quantum computation \cite{KnillMilburn:01,KokMilburn:07}, and whose linewidths are significantly narrower than those available from conventional SPDC sources, ranging from a few GHz far into the MHz range \cite{SaglamyurekTittel:11,Ranvcic:Sellars18,AskaraniTittel:19}. Similarly narrowband SPDC photons are otherwise
only achieved using complex cavity-based approaches \cite{ChuuHarris:12,WuChuu:17}.

Moreover, the wavelength of the counter-propagating photon is almost invariant to the wavelength of the pump, resulting in a phase-matching function that mimics the condition of asymmetric group-velocity-matching \cite{URenRaymer:06}, but without the dependence on material dispersion or waveguide dispersion engineering. This can be utilized to generate highly pure photons in analogy to the corresponding approach using co-propagating photons \cite{URenRaymer:06}. An illustration of these unusual properties is depicted in Fig. \ref{fig:BackwardWaveSPDC}.

Although backward-wave generation was first proposed more than 50 years ago \cite{Harris:66}, it was not realized experimentally until 2007 in a mirrorless parametric oscillator \cite{CanaliasPasiskevicius:07}, operating in the stimulated regime above the SPDC threshold, due to the difficulty associated with fabricating the required ultra-short poling periods. Since then, there have been several theoretical works establishing the underlying quantum-theory of the technique \cite{SuharaOhno:10,GattiBrambilla:15,CortiGatti:16} and its applicability for pure-photon \cite{ChristSilberhorn:09,GattiBrambilla:18} and narrow-band \cite{ChuuHarris:11} generation. Experimental demonstrations of backward-wave SPDC have been reported as a proof-of-principle for 5th-order phase-matching \cite{LatypovKalachev:17,LuoSilberhorn:20} and to generate pure \cite{LiuZhu:21} and narrowband \cite{LiuZhu:21(2)}, polarization-entangled photons, using 3rd-order phase-matching. First-order backward-wave SPDC was reported only recently, using sub-micron poling periods in bulk KTP \cite{KuoCanalias:23}.

Despite its interesting properties, the applicability of backward-wave SPDC is still limited due to the difficulty associated with fabricating the required ultra-short poling periods. However, the progress in the fabrication of sub-micrometer gratings has been steady and ongoing, and once established, we expect this technique to find much broader application. 

\bigskip

\begin{figure}
    \centering
    \includegraphics[width=1\columnwidth]{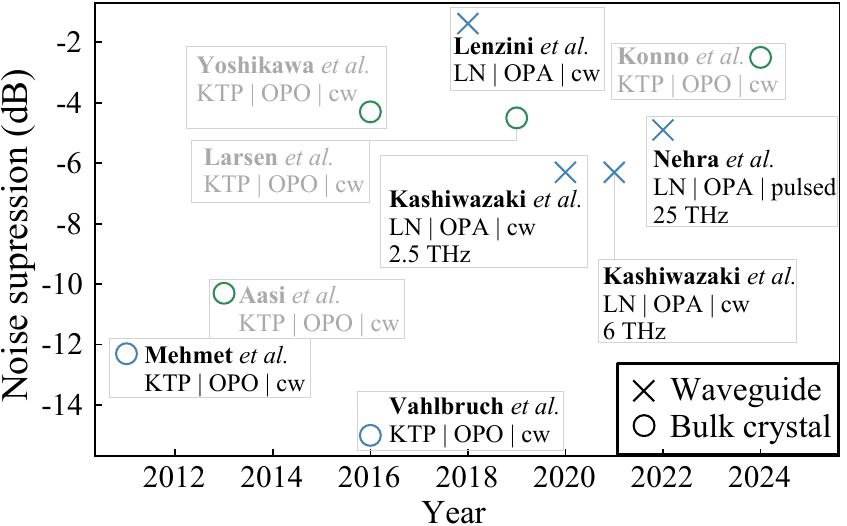}
    \caption{\textbf{Squeezed state generation} A selection of experiments demonstrating the generation (blue) and application (green) of squeezed states in recent history; featuring Mehmet \textit{et al.} \cite{MehmetSchnabel:11}, Aasi \textit{et al.} \cite{AasiZweizig:13}, Yoshikawa \textit{et al.} \cite{YoshikawaFurusawa:16}, Vahlbruch \textit{et al.} \cite{VahlbruchSchnabel:16}, Lenzini \textit{et al.} \cite{LenziniLobino:18}, Larsen \textit{et al.} \cite{LarsenAnderson:19}, Kashiwazaki \textit{et al.} \cite{KashiwazakiFurusawa:20,KashiwazakiFurusawa:21}, Nehra \textit{et al.} \cite{NehraMarandi:22} and Konno \textit{et al.} \cite{KonnoFurusawa:24}. Implementation in waveguides and bulk-crystals are marked with crosses and circles respectively. We highlight the method of generation for each experiment, differentiating between the use of optical parametric amplifiers (OPAs) and optical parametric oscillators (OPOs) operated with continuous-wave (cw) or pulsed lasers, as well as the employed crystals. Where applicable, we further list the bandwidths of broadband squeezed vacuum, detected using all-optical measurement OPAs. We note the recent focus on integrated implementation and the discrepancy between state-of-the-art sources of squeezed states and those currently employed in milestone experiments.}
    \label{fig:SqueezedStateGeneration}
\end{figure} 

\noindent
\textbf{Squeezed state generation}\\
Squeezed states constitute the fundamental quantum resource for continuous-variable based optical quantum computing \cite{AsavanantFurusawa:22,Pfister:19} and represent essential components in quantum sensing, where they are used to reduce noise limiting the sensitivity and resolution of advanced interferometers \cite{YuMavalvala:20} and microscopes \cite{CasacioBowen:21}.  

Squeezed states can be generated in high-gain (spontaneous) parametric down-conversion, operated above the threshold of single photon-pair generation, forming so-called optical parametric amplifiers (OPAs) \cite{WuWu:86}. This approach, featuring strictly periodically poled KTP and LN crystals has come to represent the go-to method of generating squeezed states across the full range of applications.

As these applications frequently utilize squeezed states with notably different characteristics, often generated in resonantly confined OPAs (forming so-called optical parametric oscillators (OPOs)), which have, at a fundamental level, remained unchanged over the last decade, a complete review is not within the scope of this paper. Nonetheless, due to the prominent role of squeezed states in photonic quantum technologies and the potential of adapting the until now strictly periodically structured crystals to more complex domain engineering techniques, we present an overview of key experiments demonstrating or featuring squeezed state generation.

The to-date strongest squeezing in optical systems was demonstrated using an OPO, achieving squeezing levels of $\sim 15 \mathrm{dB}$ \cite{VahlbruchSchnabel:16}. While OPOs represent the most common source of squeezed states \cite{VahlbruchSchnabel:16,AsavanantFurusawa:19,ChenPfister:14,LarsenAnderson:19,YokoyamaFurusawa:13,YoshikawaFurusawa:16,RoslundTreps:14,KonnoFurusawa:24}, squeezing can alternatively be generated using single-pass OPAs , generally relying on waveguide implementation or pulsed pumping to compensate for the lack of resonant enhancement \cite{KashiwazakiFurusawa:20,KashiwazakiFurusawa:21,LenziniLobino:18,NehraMarandi:22}. OPAs can further be employed to enable broadband, all-optical measurement of squeezed states \cite{InoueFurusawa:23,ShakedPeer:18,TakanashiFurusawa:20}. We present a non-exhaustive overview of experiments generating state-of-the-art squeezed states in waveguides and bulk-crystals together with milestone experiments demonstrating their application in Fig. \ref{fig:SqueezedStateGeneration}.

Although, so far, the crystals used for squeezed state generation have been structured in a strictly periodic fashion, we see considerable potential for the inclusion of aperiodic designs. Recent years have seen significant research effort committed toward the development of OPA based squeezers with the aim of generating broadband squeezed light. Large-bandwidth, few-cycle squeezed states are essential resources for high clock-rate signal processing in optical quantum computers \cite{TakedaFurusawa:19}, and enable the definition of few-cycle temporal bins compatible with short, chip-scale delay lines for time-multiplexed continuous variable quantum information processing \cite{AsavanantFurusawa:19,NehraMarandi:22}. Adapting OPAs to include chirped crystals, structured akin those used for the generation of broadband biphotons, could greatly improve the availability and quality of such few-cycle squeezed states \cite{HoroshkoKolobov:13,HoroshkoKolobov:17}. 

Aperiodically structured crystals may additionally be employed to directly tailor the spectrum of the generated squeezed states, which may be used to generate frequency-bin cluster states \cite{DragoBranzcyk:22,HurvitzArie:23} for frequency-domain quantum information processing. Few- or single-cycle squeezed states and broadband OPAs may further be used for fundamental investigations \cite{KizmannBurkard:19} and the all-optical measurement of squeezed light \cite{ShakedPeer:18}.

\bigskip

\begin{figure*}
    \centering
    \includegraphics[width=2\columnwidth]{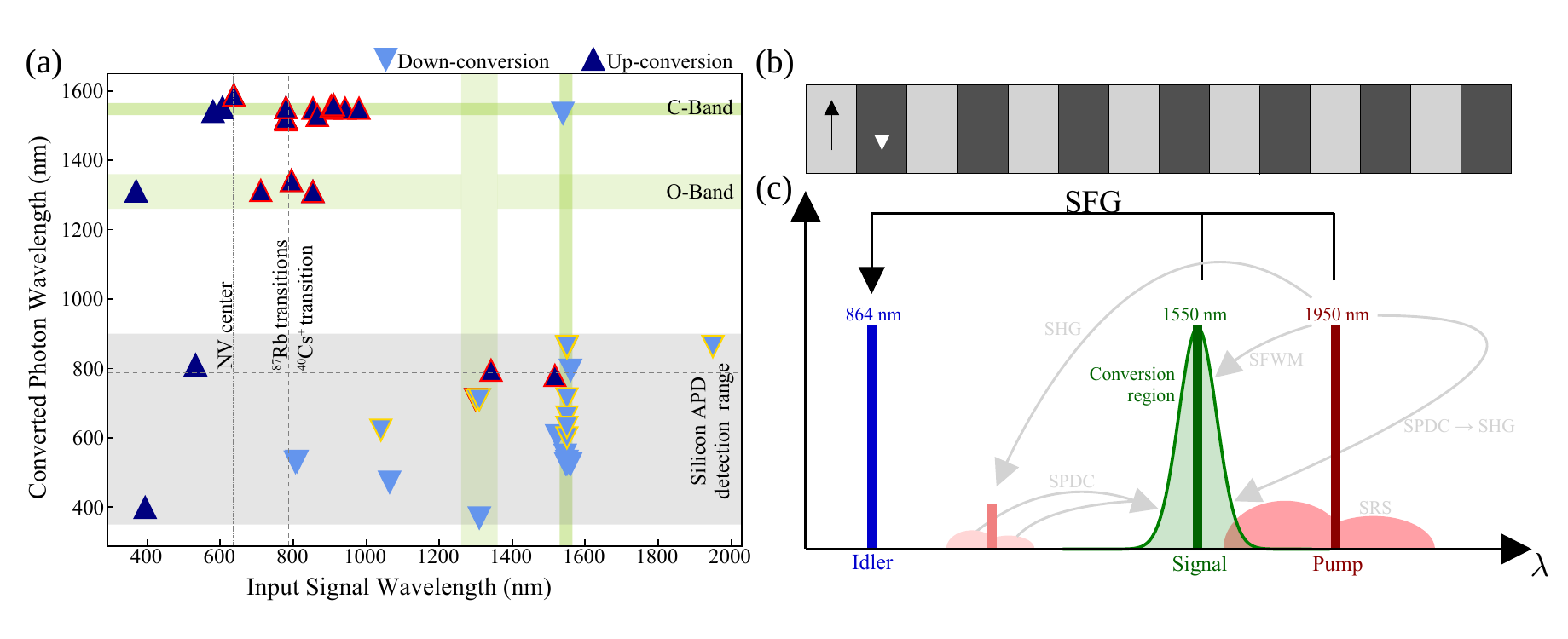}
    \caption{\textbf{Quantum frequency conversion} (\textbf{a}) A collection of the different QFC demonstrations to date. Fiber telecommunication O- and C-bands and the typical range of a visible-range silicon avalanche photon diode are highlighted. The transitions of $\mathrm{{}^{87}Rb}$ atoms, $\mathrm{{}^{40}Cs^{+}}$ ions and the NV center are indicated. Markers with red outlines correspond to quantum interface demonstrations, markers outlined in yellow to demonstrations of single-photon detectors. (\textbf{b}) Schematic of the periodic poling structure generally used to implement QFC. (\textbf{c}) Schematic illustration of the noise processes affecting QFC, including spontaneous Raman scattering (SRS) and parasitic (cascaded) three- and four-wave-mixing processes for an illustrative SFG process (including second harmonic generation (SHG), spontaneous four-wave-mixing (SFWM) and cascaded SPDC and SHG). To reduce noise-pollution, the pump is generally located outside the signal and the converted photon. Noise photons generated by the strong pump can pollute the process either directly at the wavelength of the generated idler photon, or by being converted along with the signal. Noise photons are to be removed for optimal quantum state transfer via process design or filtering.}
    \label{fig:QuantumFrequencyConversion}
\end{figure*} 

\noindent
\textbf{Quantum frequency conversion}\\
Relying on a signal photon to be incident along with the pump, sum- and difference-frequency-generation, are unsuitable for the generation of new quantum states. Both processes, however, directly offer themselves for transduction of the quantum state carried by the incident signal photon from one frequency mode to another. This so-called quantum frequency conversion represents an essential capability for the efficient interfacing of the different components constituting a quantum network. Different sources of optical quantum states, whether based on nonlinear interaction or single emitters, for instance, often operate in wavelength bands very different, both in terms of the central frequency and bandwidth, from those of detectors, memories, or repeaters. Furthermore, long-range, low-loss transmission via optical fiber networks requires operation at telecommunication wavelengths, and transmission in free-space calls for careful selection of wavelengths to account for atmospheric absorption and noise.

QFC based on sum- and difference frequency generation \cite{Kumar:90,HuangKumar:92} can, in principle, be noise-free and without uncontrolled changes to state properties other than the frequency while preserving the number statistics of the converted photon. Over the last decades, QFC has been subject of steady research and was implemented for a wide range of different transitions, an overview of which is depicted in Fig. \ref{fig:QuantumFrequencyConversion}a. QFC was demonstrated for coherent and squeezed states \cite{KerdoncuffLassen:21,LiuLi:15,BauneSchnabel:15,KongLi:14,VollmerSchnabel:14} as well as single photons \cite{WangPan:23,FisherLobino:21,ZhengPan:20,LiuShi:20,MaringRiedmatten:18,RutzSilberhorn:17,AllgaierSilberhorn:17,ZhouShi:17,KastureLobino:16,IkutaImoto:14,Takesue:08} while preserving entanglement in polarization \cite{KaiserTanzilli:19,DonohueResch:14,RamelowZeilinger:12,XiangChen:18}, time-energy \cite{LenhardEschner:17,IkutaInoto:13,IkutaImoto:11,TanzilliZbinden:05} and orbital-angular-momentum \cite{ZhouGuo:16,ZhouGuo:16(2)} degrees-of-freedom. Furthermore, a number of studies investigating the noise \cite{StrassmannAfzelius:19,PelcFejer:10,GuoWang:22,KuoTang:13,PelcFejer:11}, introduced by spontaneous Raman scattering and parasitic SPDC/SFG/DFG processes, have been presented.

QFC has been demonstrated in a number of explicit applications, a short overview of which is presented below:

\noindent
\textit{\textbf{Quantum interfaces:}}\\
The distribution of entanglement between distant nodes is an essential requirement for quantum networks, to be used in long-range quantum communication or distributed quantum computing. Implementations relying on the distribution of photons carrying entanglement with a local element are highly promising but generally rely on transduction into frequency bands suitable for efficient, low-loss transmission through optical fibers. To this end, QFC was used to establish entanglement between telecommunication-band photons and single-atoms \cite{VanLeentWinfurter:20}, atomic-ensembles \cite{IkutaImoto:18,AlbrechtRiedmatten:14}, diamond spin-qubits \cite{DreauHanson:18} and trapped-ions \cite{KrutyanskiyLanyon:19,BockEschner:18,WalkerKeller:18}, respectively. Moreover, entanglement between distant atoms \cite{VanLeentWinfurter:22}, atomic ensembles \cite{LuoPan:22,YuPan:20} and solid state-defects \cite{KnautLukin:24} as well as quantum state transfer between an atomic ensemble and a rare-earth-ion-doped crystal \cite{MaringRiedmatten:17} was demonstrated.

QFC has further been used to facilitate compatibility between single-photon sources operating outside the telecommunication band, like e.g. quantum dots \cite{MorrisonGerardot:21,WeberMichler:19,KambsBecher:16,AtesSrinivasan:12,ZaskeBecher:12,deGreveYamamoto:12,PelcFejer:12(1),RakherSrinivasan:11,RakherSrinivasan:10}, and existing fiber-communication technology.

\noindent
\textit{\textbf{Single-photon detectors:}}\\
While infrared photons offer optimal, long-range transmission, single-photon detection in the infrared band requires either noisy, inefficient InGaAs avalanche photon diodes or expensive superconducting single-photon detectors, relying on cryogenic cooling. To address this issue, QFC can be used to convert infrared photons into the (near) visible band, where cheap and efficient silicon avalanche photon diodes are available for detection \cite{MaTang:12}. This has been demonstrated to construct telecom-range single-photon detectors \cite{WangPan:23,ZhengPan:20,MaPan:18,ShentuPan:13,PelcFejer:12(2),HuangZeng:11,PelcFejer:11,MaTang:09,ThewGisin:08,HonjoInoue:07,ThewGisin:06,LangrockTakesue:05,AlbotaWong:04,RoussevFejer:04} and to establish quantum key distributions over long distances in fiber \cite{LiuPan:13,XuTang:07,DiamantiYamamoto:06} and free-space \cite{LiaoPan:17}.

QFC can further be used to enable mid-infrared imaging with single-photon resolution \cite{HuangZeng:22,WangZeng:21,DamPederson:12}, as well as spectroscopy \cite{ZhengZeng:23,ShangguanPan:16} and lidar \cite{WidarssonLaurell:22} applications.

\noindent
\textit{\textbf{Other applications:}} \\ \indent
\textit{Color erasure:} Indistinguishability is a key requirement for quantum interference, and as a result, photons located in distinct frequency bands do not interfere with one another. However, using QFC, frequency distinguishability can be erased while maintaining its Fourier-transform limited characteristics \cite{Takesue:08}. Moreover, QFC can be used to realize a frequency-domain beam-splitter, which has been used to demonstrate Hong-Ou-Mandel interference between photons of different wavelengths \cite{KobayashiImoto:16,XiangChen:18} and a frequency-domain Mach-Zehnder interferometer \cite{kobayashiImoto:17}. Detectors incorporating QFC have been used to retroactively recover interference from conventional interferometry utilizing different frequencies \cite{QuPan:19}, and to augment conventional Hong-Ou-Mandel interference \cite{YangShi:22}. Erasing the color-information of photons greatly relaxes the wavelength compatibility demands of quantum networks.

\textit{Bell-state measurement:} Bell-state measurements represent a key component for quantum information protocols based on entanglement swapping and quantum teleportation. However, using linear optics, only partial Bell-state measurements, resolving two of the four states, is possible \cite{SunPan:17}. Based on the selective, narrowband interaction between two incident photons in a three-wave-mixing process, SFG can be used to overcome this limitation. Measurement and distinguishing of the four Bell-states was first demonstrated in the polarization degree-of-freedom, to realize quantum state teleportation \cite{KimShih:01}. The approach was then adapted for entanglement swapping \cite{SangouardZbinden:11}, where it acts as a nonlinear filter to remove the multi-pair component complicating approaches based on linear optics. Recently, sum-frequency-based Bell-state detection has been employed to demonstrate multi-user entanglement swapping \cite{LiChen:19} and a quantum secure direct communication network \cite{QiChen:21}.

\textit{Time-frequency mode control:} Time-frequency modes form a complete framework for photonic quantum information science \cite{BrechtRaymer15}. This is based on the fact, that any single-photon TFM mode can be expressed in a basis of mutually orthogonal modes, which can be addressed selectively using a SFG process (or, in principle, a DFG process), if the pump-field is shaped to match the TFM \cite{ReddyRottwitt:13}. This was found to occur optimally in a specific group-velocity matching configuration, to form a so-called quantum pulse gate \cite{EcksteinSilberhorn:11}. The concept has since been developed to allow for the shaping of TFM \cite{BrechtSilberhorn:11} achieving near unit mode selectivity \cite{ReddyMcKinstrie:14,ReddyRaymer:15}, and was demonstrated using coherent-states \cite{BrechtSilberhorn:14,ManurkarKanter:16,KowligyHuang:14,ReddyRaymer:17,ShahverdiHuang:17,ReddyRaymer:18} as well as genuine single-photon quantum states \cite{AnsariBrecht:18}. Quantum pulse gates have further been implemented as single-photon subtractors \cite{RaTreps:17}, to measure unknown superpositions \cite{AnsariSilberhorn:18(2)}, and recently, to realize a multi-output quantum pulse gate, based on an aperiodic nonlinearity profile \cite{SerinoSilberhorn:23}. 

\textit{Ultrafast correlator:} The SFG (DFG) process can be used as an ultrafast correlator between either two incident photons or a single photon and a pump pulse. This allows to resolve ultrafast temporal dynamics, and can be employed, for instance, to directly observe the two-photon wavefunction. This has been used to characterize broadband entangled photons \cite{DayanPeer:05,SensarnHarris:10} and to measure the wavefunction of shaped photon-pairs \cite{PeerSilberberg:05,ZahFeurer08}.

QFC already represents a well-established technology in both waveguides and bulk crystals, and has found application in wide range of technologies. Its efficiency is limited predominantly by fabrication imperfections and limits on the pump power. Device performance can suffer from noise-pollution as a consequence of secondary effects resulting from a strong pump, as illustrated in Fig. \ref{fig:QuantumFrequencyConversion}c, but can be reduced to very low levels by appropriate process design and filtering. Performance both in terms of efficiency and low-noise operation has been steadily increasing. To date, QFC almost exclusively relies on the periodically-poled crystals indicated in Fig. \ref{fig:QuantumFrequencyConversion}b, but we see definite potential for the incorporation of more advanced domain engineering techniques. Adiabatic techniques \cite{SuchowskiArie:14}, for instance, can be applied to increase the acceptance bandwidth of the conversion, and multi-period gratings could be utilized to realize devices operating at multiple input/output bands simultaneously.  

\bigskip

\noindent
\textbf{Further Applications}\\
At last, we would like to note applications relying on SPDC or single-photon sum- or difference-frequency generation developed over the last two decades, that did not fit any of the categories above. We note that this list may very well not be exhaustive. 

(\textit{1}) Combining frequency-chirped photons and pump pulses allows control over the converted photon's waveform in time-frequency space \cite{KielpinskiWiseman:11,DonohueResch:15}, which has been used to generate narrowband photons \cite{LavoieResch:13} and resolve narrow time-bin encodings \cite{DonohueResch:13}.

(\textit{2}) Recently, adaptive, aperiodic poling was utilized to eliminate the impact of nano-scale inhomogeneities in nanophotonic LN waveguides, which limit the nonlinear efficiency \cite{ChenFan:24}. This was demonstrated to drastically increase the efficiency of second-harmonic-generation to an order of magnitude above that of periodically poled crystals, but is equally applicable for quantum three-wave-mixing processes.

\bigskip

\noindent
\textbf{Summary and Outlook}\\
Over the last decades, the field of quantum technologies has increasingly shifted from the development of fundamental concepts and proof-of-principle experiments to attempts of real-world implementation, which has come to demand increasingly complex resources. Photonic quantum technologies constitute a promising platform in a multitude of different applications and currently represent the only option for realizing distributed networks. Satisfying the often very different demands of these applications will require high amounts of platform flexibility and compatibility.

Here, we reviewed an engineering approach promising such flexibility, based on manipulating the photon generation process originating from the nonlinear interaction with the surrounding crystal. Over the last decades, this approach has been developed to realize a number of vastly different goals. Most notably, arbitrary shaping of bi-photon spectra allowed direct control of their entanglement properties, multi-process phase-matching functions enabled control of the bi-photons' spectral and polarization characteristics, and single photon frequency conversion processes were used to interface different quantum systems.

While crystals structured with simple periodic patterns already represent integral components in many cutting-edge experiments, more sophisticated designs, like the ones described here, have seen only limited application. In the coming years, we envision progressive integration of techniques featuring aperiodic structures, fully exploiting the potential offered by the domain engineering approach, into large, front-line experiments.  

Although not discussed in this review, the development of the domain engineering approach has been paralleled by the ongoing effort of developing integrated photonic platforms across a number of different materials. Realizing photon sources or interconnects in an integrated manner using nonlinear waveguides offers orders-of-magnitude increased nonlinear efficiency and unrivaled stability and scalability, with respect to implementation in bulk optics.

The production of high-quality waveguide structures and the development of poling techniques applicable to thin-film wafers continue to represent large, ongoing areas of research and development, and the uniformity of the waveguides' geometry as well as the quality of the inverted domains have seen significant improvement in recent years. The minimum available inverted domain size, too, has seen major improvement, enabling more compact designs and more freedom in the design-process. We envision a continuous shift from bulk-optic to integrated-optic implementation, driven by the demands associated with demonstrating optical quantum technologies in real-world environments.

The techniques used to structure the material nonlinearity, during or after crystal growth, are currently available only in a limited number of materials, and several of the applications discussed in this review further rely on particular, material-specific dispersion properties. Both of these factors can significantly limit the applicability of the respective techniques. Bypassing these limitations by engineering or adapting for material dispersion, developing poling techniques in new materials, or combining material-platforms in a hybrid fashion, would represent major breakthroughs.

We see considerable potential in incorporating nonlinear domain engineering techniques with photonic circuits. Including resonant confinement to modify the nonlinear interaction or cascading multiple nonlinear elements to form nonlinear interferometers, for instance, hold particular promise. We further anticipate applications featuring combinations of the different techniques, or the adaption of classical domain engineering schemes relying on e.g. cascaded processes or adiabatic conversion, to realize more capable, compact and robust devices. 

Of course, the most interesting future development is likely to stem from adapting the technique for wholly new purposes, and we would like to encourage researchers to further explore this approach, which we believe to still be far from reaching its potential.

\bigskip
\noindent
\textbf{Acknowledgments}\\
AP acknowledges an RMIT University Vice-Chancellor’s Senior Research Fellowship a Google Faculty Research Award and support by the Australian Government through the Australian Research Council under the Centre of Excellence scheme, (No: CE170100012). 

\bigskip
\noindent
\textbf{Conflict of Interest Statement}\\
The authors have no conflicts to disclose.

\bibliography{References_V2}

\clearpage
\setcounter{figure}{0}
\makeatletter 
\renewcommand{\thefigure}{S\@arabic\c@Fig}
\makeatother

\setcounter{equation}{0}
\makeatletter 
\renewcommand{\theequation}{S\@arabic\c@Eq}
\makeatother

\setcounter{table}{0}
\makeatletter 
\renewcommand{\thetable}{S\@arabic\c@table}
\makeatother

\end{document}